\documentclass{article}
\usepackage{ijcai19}

\usepackage{soul}
\usepackage{url}
\usepackage[hidelinks]{hyperref}
\usepackage[utf8]{inputenc}
\usepackage{times}
\usepackage{algorithm}
\usepackage{algorithmic}
\usepackage{xcolor}
\usepackage[cmex10]{amsmath}
\usepackage{graphicx}
\usepackage{multirow}
\usepackage{amsfonts}
\usepackage{multirow}
\usepackage{booktabs}
\usepackage{subfigure}
\usepackage{amsmath,amsthm}
\usepackage{mathtools}

\title{Wasserstein Autoencoders for Collaborative Filtering}

\author{
Jingbin Zhong$^1$\and
Xiaofeng Zhang$^2$\\
\affiliations
$^1$Harbin Institute of Technology, Shenzhen\\
$^2$Harbin Institute of Technology, Shenzhen\\
\emails
zhongjingbin@stu.hit.edu.com,
zhangxiaofeng@hit.edu.com
}

\begin{document}
\maketitle

\begin{abstract}
The recommender systems have long been studied in the literature. The collaborative filtering is one of the widely adopted recommendation techniques which is usually applied on the explicit data, e.g., rating scores. However, the implicit data, e.g., click data, is believed to be able to discover user's latent preferences. Consequently, a number of research attempts have been made towards this issue. In this paper, we propose to adapt the Wasserstein autoencoders for this collaborative filtering task. Particularly, we propose the new loss function by introducing an $L_1$ regularization term to learn a sparse low-rank representation form for the latent variables. Then, we carefully design (1) the new cost function to minimize the data reconstruction error, and (2) the suitable distance metrics for the calculation of KL divergence between the learned distribution of latent variables and the underlying true data distribution. Rigorous experiments have been evaluated on three widely adopted datasets. Both the state-of-the-art approaches, e.g., Mult-VAE and Mult-DAE, and the baseline models are evaluated and the promising experimental results have demonstrated that the proposed approach is superior to the compared approaches with respect to criteria $Recall@R$ and $NDCG@R$. 
\end{abstract}

\section{Introduction}\label{sec:Intro}

The recommender systems have long been studied in the literature and various kinds of approaches have been proposed~\cite{ricci2015recommender,Georgiev2013A,Shi2014Collaborative}. Among them, collaborative filtering (CF) is one of the most important recommendation techniques~\cite{koren2015advances,herlocker2017algorithmic}. The CF recommends items either by user-based algorithms or item-based algorithms employed on the explicit user data (e.g.,  ratings)~\cite{davidson2010youtube}. However, the implicit user data, e.g., click data and browsing historical data~\cite{li2017classification}, are believed to contain user's potential preferences which attracts more and more research efforts with a focus on predicting items from the implicit data \cite{joachims2017accurately}. 

The implicit data are generally sparse and noisy as they are usually collected from the real-world applications. This mainly hinders the linear CF models, such as matrix factorization (MF) based approaches \cite{koren2009matrix}, to achieve a comparably good prediction accuracy due to their limited data representation ability. Consequently, a number of neural collaborative filtering algorithms are proposed to generalize linear latent-factor models to endow them with non-linear data representation and the prediction \cite{kuchaiev2017training}.
For instance, 
the proposed Mult-VAE \cite{liang2018variational} extends variational autoencoders for collaborative filtering task. It assumes the distribution of latent variables could be estimated from the implicit data. Then, the latent variables are sampled from the estimated distribution. At last, the reconstructed data can be decoded from the latent variables through the multilayer neural network.




Although, the VAE based approaches have achieved the state-of-the-art recommendation performance, it forces $P(Z|X)$, the distribution of latent variable $Z$ learned from input data $X$, to approximate the assumed ground-truth distribution of latent variable $P(Z)$ for all input data. This might increase the reconstruction error as it maps multiple input data items to the same output data item but not different data items. Alternatively, the Wasserstein autoencoders (WAE) \cite{tolstikhin2017wasserstein} is proposed for this mathematical problem. However, it is not a trivial task to adapt the WAE for the collaborative filtering task as the implicit data is generally a large-scale sparse matrix and technical efforts must be made for this issue. Motivated by this, we carefully adapt the Wasserstein autoencoders approach for collaborative filtering task (hereinafter ``aWAE''). To the best of our knowledge, this is the first attempt to adapt WAE to collaborative filtering problem. In the proposed aWAE, the new loss function is proposed by introducing an $L_1$ regularization term to allow a sparse low-rank representation form for the generated latent variables. To optimize the $L_1$ regularization loss, the standard ADMM \cite{boyd2011alternating} algorithm is employed which separately learns the solutions to the constrained optimization problems. The rest model parameters of the proposed aWAE are separately learned via the variational inference learning \cite{graves2011practical}. The contributions of this paper can be summarized as follows. 


\begin{itemize}
\item  To the best of our knowledge, this is the first attempt to adapt Wasserstein autoencoders (aWAE) for collaborative filtering task. Technically, we design the overall framework as well as its components to facilitate the recommendation from the sparse implicit data. Particularly, we propose a new objective function by introducing an $L_1$ regularization term to allow the sparse low-rank representation for the generated latent variables. 

\item We design a new cost function to minimize the data reconstruction error and propose a sample mean-variance method (SMV) which is suitable for the calculation of the KL divergence. We also propose a modified variational inference learning algorithm for the learning of the parameters of deep latent networks and the corresponding ADMM updating rules are also reformulated to resolve the the constrained optimization problem separately.


\item Rigorous experiments have been performed on three real-world datasets, i.e., ML-20M, Netflix and LASTFM. Several baseline models as well as the state-of-the-art approaches are evaluated for the performance comparison which are Mult-DAE \cite{liang2018variational}, Mult-VAE \cite{liang2018variational}, CDAE \cite{wu2016collaborative} and Slim \cite{ning2011slim}. The experimental results have demonstrated the superiority of the proposed aWAE with respect to two widely adopted evaluation criteria, i.e., $Recall@R$ and $NDCG@R$.  
\end{itemize}

\section{Related Works}\label{sec:Related}

Conventionally, the recommender systems are designed to predict ratings which might be assigned by a target user on unscored items based on the actual ratings of a group of users with common tastes \cite{Georgiev2013A}.
There exist a good number of recommendation techniques \cite{ricci2015recommender,Georgiev2013A,Shi2014Collaborative}. Among them, the collaborative filtering (CF) \cite{Yi2017Collaborative} based approaches play an important role. The CF recommends items either by user-based algorithms or item-based algorithms employed on the explicit user data (e.g., ratings)~\cite{davidson2010youtube}. In \cite{mnih2008probabilistic}, the probabilistic matrix factorization (PMF) approach is proposed which is proved to be able to cope with large and sparse training data with a superior performance. In this paper, the proposed PMF tries to find an appropriate low rank representation for the association between a large user matrix and item matrix. These low rank representations are believed to well interpret users' preferences. In fact, a large body of CF approaches are essentially linear models which are not robust enough to model the implicit data. 

The implicit data includes such as click data and browsing historical data~\cite{li2017classification}. These implicit data can be used to discover user's potential interests in particular items and thus attract more and more research efforts \cite{joachims2017accurately}. The natural choice to analyze such huge amount of implicit data is to employ deep neural network models  \cite{Xu2016Tag}. In \cite{makhzani2016adversarial}, the autoencoder based approach is proposed which assumes that the low-dimensional latent variables are responsible for the generation of high-dimensional click data and then several autoencoder based approaches have been proposed for the CF problems \cite{kingma2013auto,DBLP:xin:asdae}. In \cite{chen2018collective}, a collective variational autoencoder is proposed to recommend top-N items by using the auxiliary information. In this approach, both users' side information and item's side information are modeled using autoencoder and the latent variables are assumed to follow a Gaussian distribution. Then, the output is binarized to capture the implicit feedback. The recent proposed Mult-VAE \cite{liang2018variational} first assumes the implicit data follows a multi-nomial distribution, and then the latent variables are encoded from a multilayer deep latent neural network. After estimating the statistics of the latent variable distribution, the latent variables are sampled from this learned distribution. At last, the reconstructed data is decoded through the nonlinear mapping from the sampled latent variables. The proposed loss function is to minimize the reconstruction error between input implicit data and the output data, which already achieves the state-of-the-art prediction results on implicit feedback data. 
However, one problematic issue in VAE based approaches is that the distributions of latent variables overlap a lot which might increase the reconstruction error. Motivated by the Wasserstein autoencoders approach (WAE) \cite{tolstikhin2017wasserstein}, we propose this work to investigate how to extend Wasserstein autoencoders for collaborative filtering task.




\section{The Proposed Approach}\label{sec:Approach}
\subsection{Problem Formulation}
Let $X ^{N\times M}$ denote the implicit  
data like click data where $N,M$ respectively denote the number of users and items, $x_{i} = [x_{i1}, ..., x_{iM}]^{T} \in X$
and $x_{ij}=1$ represents that the $i$-th user clicks the $j$-th item and 0 otherwise. Generally, $X$ could be binarized from users' rating matrix by assigning 1 to the entries whose rating score is greater than a predefined threshold, and $X'$ is the reconstruction of $X$. Similar to \cite{liang2018variational}, we also assume that the click data $X$ obeys a multinomial distribution, written as
\begin{equation}\label{eq:XDist}
x_i \sim Mult(M_i,\delta(.)),
\end{equation}
where $M_i = \sum \limits_{j} x_{ij}$ is the total number of clicks by user $u_i$, $\delta(\cdot)$ outputs the corresponding probability for each click number in $[0, M_i]$. Generally, $\delta(\cdot)$ is assumed to be a softmax function to guarantee the summation of probability to be 1. 



\begin{figure}[!ht]
	\centering
	\includegraphics[width=3.5in,height=1.6in]{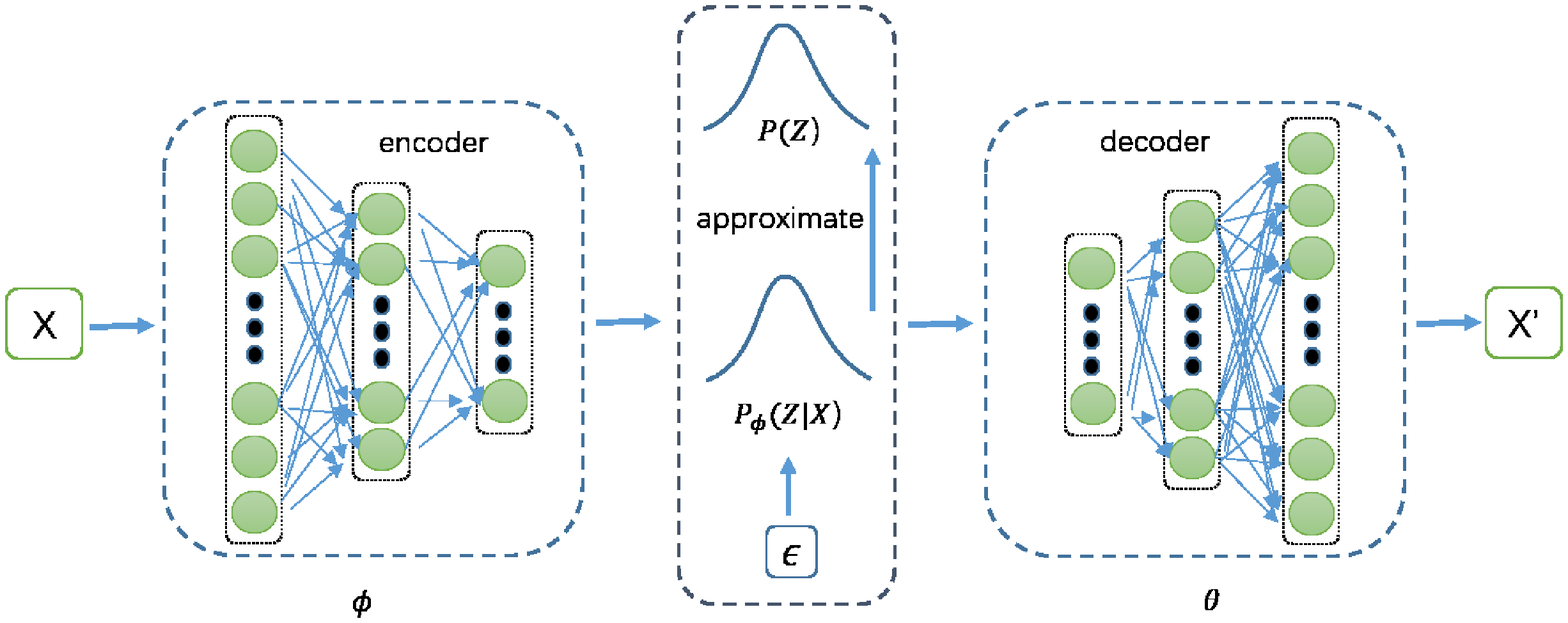}
	\caption{Framework of the proposed approach. }
	\label{fig:abVSl}
\end{figure}

\subsection{The Proposed aWAE}

The framework of the proposed adapted Wasserstein Autoencode (aWAE) for collaborative filtering task is depicted in Figure \ref{fig:abVSl}. The main steps of the proposed aWAE are illustrated as follows. First, the implicit data $X$ is first encoded through an encoder component, highlighted in blue dashed rectangle, to directly generate the latent variables $Z$. The learned distribution $P(Z|X)$ from the generated latent variables $Z$, during the model learning process, is forced to approximate the underlying true distribution $P(Z)$ of latent variables. To avoid the over-fitting issue, a certain level of random noise $\epsilon$ is naturally applied to $Z$. Second, the twisted $Z$ (with the addition of random noises) is decoded through the decoder component, highlighted in the red dashed rectangle, to reconstruct $X'$ which is forced to resemble $X$, written as $||X'-X||<\eta$, where $\eta$ is a small enough positive number. 

Note that the proposed aWAE consists of two important components, i.e., encoder and decoder components. For the encoder component, it is natural to employ a multilayer neural network, denoted as $g_\phi(\cdot)$ parameterized by $\phi$, to acquire a robust nonlinear data transformation capability and it outputs $g_{\phi}(X)$ which is treated as the observed latent variables. Thus, we have $Z=g_{\phi}(X)$. Alternative to the VAE approach, the latent variables $Z$ are directly transformed from the implicit data $X$ through a dimension reduction process (achieved by $g_\phi(\cdot)$). According to \cite{tolstikhin2017wasserstein}, the observed latent variables $Z$ are usually assumed to obey a Gaussian distribution, written as $z_{i} \sim \mathcal{N}(0,1)$. Note that for many real-world applications, $X$ is generally large and sparse and only a few latent variables would account for the eventual recommendations. Therefore, the dimension of $Z$ should be restricted to a small number in the empirical studies. As for the decoder component, it maps a low-dimensional $Z$, through a multilayer neural network $f_\theta(\cdot)$, parameterized by $\theta$, to the high dimensional $X'$ and we have $X'=f_{\theta}(Z)$. Obviously, a set of neural network parameters, e.g., $\phi$ and $\theta$, should be learned during the model learning process.

\subsection{Variational Inference Learning}

To learn the proposed aWAE, the variational information learning \cite{graves2011practical} is a natural choice. Accordingly, the penalized Evidence Lower Bound (ELBO) of the original Wasserstein autoencoders (refer to \cite{tolstikhin2017wasserstein}) is given as  
\begin{eqnarray}\label{eq:OrgWAE}
&L_{\beta} (x_i;\theta,\phi)=\inf\limits_{q_{\phi} \sim \mathcal{Q}} E_{P_X} E_{q_{\phi}} [c(x_i,p_{\theta}(x_i |z_i))] \nonumber \\
&+ \beta\cdot \mathcal{D}_{Z} (q_{\phi} (z_i | x_i ), p (z_i)),
\end{eqnarray}
where $\mathcal{Q}$ is any nonparametric set of probabilistic encoders, $P_X$ is multinomial prior as aforementioned, $c(x_i,p_{\theta}(x_i|z_i))$ is any measurable cost function taking two parameters $x_i$ and $p_{\theta}(x_i|z_i)$, $\mathcal{D}_Z$ could be any divergence measurement calculating the distance between two distributions $q_{\phi} (z_i | x_i )$ and $p(z_i)$. $\beta > 0$ is the parameter controlling the strength of the distance regularization term. 

As aforementioned, the dimension of $Z$ is practically required to be much smaller than that of $X$. This hints us to introduce an $L_1$ regularization term to restrict $Z$ to be as sparse as possible. To this end, we first use $S \times A$ to represent $Z$. Let $S = [s_1, s_2, ..., s_n]^T\in R^{N \times K}$ denote a low-rank sparse matrix for each latent variable $z_n \in Z$, and $A = [a_1,a_2,...,a_h] \in R^{K \times h}$ denote the coefficient matrix of $S$. Then, the following term should be considered in the overall loss function, written as 
\begin{eqnarray}\label{eq:sparse}
L_{sparse} = \lambda_1||Z - SA||^2_F + \lambda_2||S||_1,
\end{eqnarray}
where $\lambda_1$ and $\lambda_2$ are two controlling parameters to be learned. Consequently, the new ELBO problem could be formulated as follows
\begin{eqnarray}\label{eq:obj}
&L(x_i;\theta,\phi)=\inf\limits_{q_{\phi} \sim \mathcal{Q}} E_{P_X} E_{q_{\phi}} [c(x_i,p_{\theta}(x_i |z_i))] \nonumber \\
&+ \beta\cdot \mathcal{D}_{Z} (q_{\phi} (z_i | x_i ), p (z_i)) 
+ \alpha(\lambda_1||z_{i} - s_{i}A||^2_2 + \lambda_2||s_{i}||_1)
\end{eqnarray}
\subsubsection{Choosing the cost function}
Particularly, the first term in Eq.~\ref{eq:obj} is to optimize the data reconstruction error, i.e., the reconstructed $X'$ is required to be close enough to the input $X$. This is usually achieved by a carefully-chosen cost function $c(\cdot)$. As $X$ is assumed to obey a multinomial distribution, and thus it is natural for the Mult-VAE~\cite{liang2018variational} to chose the multinomial loss as its cost function, given as
\begin{equation}\label{eq:cost}
    c(x_i,p_{\theta}(x_i|z_i)) =\sum \limits_{j} x_{ij} \log \delta(f_{\theta}(z_i)).
\end{equation}

Apparently, such cost function might be problematic as it only considers the situation when $x_i\neq 0$ but ignores the situation when $x_i=0$. 
If $x_i=0$, the output of this cost function indicates that the corresponding item is never clicked by a user. However, this is not the real case as some unobserved data, although is clicked by a user, may be treated as a non-click 0 by Eq.~\ref{eq:cost}. To take into account the unobserved click data, a penalty term is applied and the proposed cost function is written as
\begin{eqnarray}
    c(x_i,p_{\theta}(x_i | z_i))=&\sum\limits_{j}x_{ij}\log \delta(f_{\theta}(z_i))+\nonumber\\ 
     & \gamma (1-x_{ij}) \log \delta(f_{\theta}(z_i)),
\end{eqnarray}
where $\delta$ is a softmax function and $\gamma$ is the weight of the penalty. In this experiments, we also evaluate other cost functions such as the missing information loss (MIL) \cite{arevalo2018missing}.
\subsubsection{Choosing distance metrics for KL divergence}
The second term of Eq.~\ref{eq:obj} is to restrict the generated $Z$ to obey the assumed prior distribution, e.g., a Gaussian distribution~\cite{zhao2017infovae}. 
This means that we should first estimate the statistics of the distribution for the generated $Z$, then we need to minimize the distance between two data distributions. The KL divergence $\mathcal{D}_Z$ is generally adopted to measure distance between different data distributions. The original WAE proposes two distance metrics to calculate this $\mathcal{D}_Z$, i.e., GAN-based $\mathcal{D}_Z$ and MMD-based $\mathcal{D}_Z$ methods. 

However these two functions are more suitable for the dense data, e.g., image data, but not the sparse data, e.g., rating data, where the distance metrics should be carefully designed. Alternatively, we propose a sample mean-variance method, called SMV method, to calculate $\mathcal{D}_Z$. Specifically, we compute the mean $\mu_i$ and the variance $\sigma_i$ for the generated $Z$ at each iteration, denoted as $z_i$. Let $J$ be the dimension of $z_i$, then the SMV could be calculated as follows
\begin{eqnarray}
\mathcal{D}_Z = \frac{J}{2}\cdot(\mu_i^2 + \sigma_i^2 -\log(\sigma_i^2) - 1)
\end{eqnarray}
The SMV method is the simplified version of method proposed in~\cite{kingma2013auto}. The original method computes vector-wise mean and variance from sample data, whereas our approach calculates a single mean and variance as WAE requires all dimensional data follows the same distribution, and thus saves a lot of computational cost. 

\subsection{The adapted ADMM algorithm}

The third term in Eq.~\cite{liang2018variational} is to allow a low-rank sparse representation of the latent variables $Z$. To resolve this $L_1$ regularization term, the alternating direction method of multipliers (\textmd{ADMM})  \cite{boyd2011alternating} algorithm could be adopted. The ADMM separates the original optimization problems into two sub optimization problems, and then optimizes two separate sub problems in an iterative manner. Suppose the parameter set $\{\phi, \theta\}$ of aWAE is already acquired, then we can fix this parameter set unchanged and update $A, S$ to satisfy following objective functions, given as
\begin{eqnarray}
     \hat{A} &=& \mathop{\arg\min}_{A} \lambda_1||Z - SA ||_F^2{\hspace{1em}} s.t.\hspace{0.5em}||a_j||^2 \leq 1, \\\label{eq:ADMMA}%
    \hat{S} &=& \mathop{\arg\min}_{S} \lambda_1 ||Z - SA ||_F^2 + \lambda_2||S||_1
\end{eqnarray}
To solve this problem via ADMM, an additional matrix $H$ is introduced and $H=A$. Thus, the corresponding new objective function is rewritten as 
\begin{eqnarray}
    & \hat{A} =\mathop{\arg\min}_{A} \lambda_1||Z - SA ||_F^2 \nonumber\\
    \nonumber\\
     &s.t. \hspace{1em}H=A, \hspace{1em}||h_j||^2 \leq 1
\end{eqnarray}
Therefore, the optimal solution $\hat{A}$ could be updated according to the following equations:
\begin{equation}\label{eq:admm update}
\left\{
\begin{aligned} 
A^{t+1} &= \mathop{\arg\min}_A ||Z-SA||_F^2 + \rho||A-H^t+U^t||_F^2\\
H^{t+1} &= \mathop{\arg\min}_H \rho ||A-H^t+U^t||_F^2 \hspace{1em} s.t. ||h_j||^2_2 \leq 1\\
U^{t+1} &= U^{t} + A^{t+1} - H^{t+1}
\end{aligned}
\right.
\end{equation}
Similarly, $S$ could be updated in the similar manner. To summarize, the parameters of the proposed aWAE are iteratively learned by minimizing the objective function in Eq.~\ref{eq:obj}. And the $L_1$ regularization term is separately updated by using the adopted ADMM algorithm. The model learning algorithm of the proposed aWAE foris illustrated in Algorithm \ref{alg:Framwork}.

\begin{algorithm}[h]
\caption{The adapted Wasserstein autoencoders (aWAE) algorithm for collaborative filtering.}
\label{alg:Framwork}
\begin{algorithmic} 
\REQUIRE ~~\\ 
    Click data $X$; $k$, $h$ (dimension) of $Z$;\\
    Regularization coefficient: $\alpha,\beta,\lambda_1,\lambda_2> 0$.\\
    Initialization: matrix $S\in R^{n \times k}$, $A\in R^{k \times h}$;\\
    Initialization: parameters $\phi$ of the encoding multilayer networks $Q_{\phi}$, and parameters $\theta$ of the decoding multiplayer network  $G_{\theta}$.\\

\ENSURE

\WHILE{($\phi, \theta$) not converged}

\STATE Sample $\{ x_1, \ldots ., x_n \}$ from the training set\\
\STATE Sample $\{ z_1, \ldots ., z_n \}$ from the prior $P_Z$\\
\STATE Sample $_{} \widetilde{z_i}$ from $Q_{\phi} (Z | x_i
)$ for i = 1,...,n\\

\STATE Fix $S$ and $A$, update $Q_{\phi}$ and $G_{\theta}$ by descending:\\
\STATE
\begin{eqnarray*}
	\begin{aligned}
		\frac{1}{n}&\overset{n}{\underset{i = 1}{\sum}} c (x_i,G_{\theta} (\widetilde{z_i})) +   \frac{\beta J}{2}(\mu_i^2+\sigma_i^2 -\log(\sigma_i^2)-1)\\
		&+\alpha \frac{1}{n} \sum^{n}_{i=1}(\lambda_1||z_i-s_i A||^2_2 + \lambda_2||s_i||_1)\\
	\end{aligned}
\end{eqnarray*}
\STATE Fix \{$\theta, \phi$\}, update $S$ and $A$ using Equation \ref{eq:admm update}. 
\ENDWHILE

\label{code:fram:isReturn}
\end{algorithmic}
\end{algorithm}

\section{Performance Evaluation}\label{sec:Exp}

In the experiments, three widely adopted benchmarck datasets, i.e., ML-20M\footnote{http://grouplens.org/datasets/movielens}, Netflix \footnote{http://www.netflixprize.com} and LASTFM \cite{Herrada2008Music} dataset, are chosen for the evaluation of model performance. Details of these datasets will be illustrated in Section \ref{sec:data}. Two state-of-the-art approaches, i.e., Mult-VAE and Mult-DAE \cite{liang2018variational}, as well as some baseline models, i.e., SLIM \cite{ning2011slim},
and CDAE \cite{wu2016collaborative}, are chosen for the model comparison. Two widely adopted evaluation criteria, e.g., $Recall@R$ and $NDCG@R$, are adopted for measuring the performance of each model. We implement both the proposed approach as well as the compared ones and report the corresponding empirical study results. The promising experimental results have demonstrated that the proposed approach is superior to both the state-of-the-art approaches as well as the baseline models with respect to the evaluation criteria. Note that the proposed aWAE is excellent in recommending only a few items, demonstrated from the experimental results, which is more meaningful for the real-world applications. 
\subsection{Datasets and Data Preprocessing}\label{sec:data}
Three widely adopted benchmarck datasets are chosen for the performance evaluation and details of each dataset are given as follows. 

\begin{itemize}
    \item \textbf{MovieLens-20M (ML-20M).} This dataset is one of the most widely adopted movie rating data set which collects users' rating scores on movie items. To process the data, we binarize the explicit ratings by keeping at least four scores and treat them as the click data (user's implicit feedback). Note that we only keep users who have scored on at least five items.
    \item \textbf{Netflix Prize (Netflix).} This data set is also a user-movie rating dataset collected from the Netflix Prize7. Similar preprocessing steps are taken to binarize the rating matrix to the implicit data matrix. 
    \item \textbf{Last.fm (LASTFM).} This dataset contains the structured records indicating whether a user is the audience of a particular artist. In the experiments, the artist with less than 50 distinct audiences will be filtered out from the dataset. Each user is restricted to follow at least 20 artists. In the converted binary-valued matrix, 1 denotes a user is the audience of an artist and 0 otherwise.
\end{itemize}

\subsection{Baseline Models}
We compare the proposed approach with the following state-of-the-art and the baseline methods. 
\begin{itemize}
    \item \textbf{Mult-DAE and Mult-VAE} \cite{liang2018variational}. These two approaches are considered as the state-of-the-art ones. They adopt variational autoencoders for colloborative filtering by assuming the implicit feedback data follows a multinomial distribution. We have implemented both approaches using  the same parameters as those in the original paper. 
    \item \textbf{Slim} \cite{ning2011slim,levy2013efficient}. Essentially, this approach is a linear model which tries to recommend items from a sparse item-to-item matrix. As the SLIM needs to grid-search the best parameters, we simply report the original results in this paper. 
    \item \textbf{Collaborative Denoising autoencoder (CDAE)} \cite{wu2016collaborative}. The CDAE extends the denoising autoencoders (DAE) by adding a latent variable. In the experiments, the size of latent variables is set to 200 as that of Mult-VAE and the proposed aWAE.  
\end{itemize}

\subsection{Evaluation Criteria}
To evaluate the model performance, two widely used evaluation metrics are adopted in the experiments which are $Recall@R$ and $DCG@R$. For criterion $Recall@R$, the top $R$ items are equally weighted and we compare the rank of predicted items with the ground truth rank, calculated as
\begin{eqnarray}
\nonumber Recall@R(u,w) = \frac{\sum_{r=1}^R I[w(r)\in I_u]}{min(M,|I_u|)},
\end{eqnarray}
where $w(r)$ denotes the item with rank $r$, $I(\cdot)$ is an indicator function, $I_u$ is the set of held-out items clicked by user $u$. In the experiments, we normalize $Recall@R$ using the minimum $R$. That is, we rank all relevant items to the top $R$ items. 

For discounted cumulative gain criterion, denoted as $DCG@R(u,w)$, it calculates the accumulated importance of all ranked items $u$. The importance of each ranked item is discounted at lower ranks, calculated as
\begin{eqnarray}
\nonumber DCG@R(u,w) = \frac{\sum_{r=1}^R 2^{I[w(r)\in I_u]}-1}{log(r+1)}.
\end{eqnarray}
The original $DCG@R(u,w)$ measures the quality of the rankings as it assigns a higher weight to the items with a higher rank. In addition, $NDCG@R(u,w)$ normalizes the original $DCG@R(u,w)$ which is adopted in our experiments. 


\subsection{Experimental Settings}

In the experiments, we follow the settings in the Mult-VAE \cite{liang2018variational} to employ 2-layer neural networks for both the encoding and the decoding multilayer neural networks. The dimension of latent variables $Z$ is empirically tuned to 200 to achieve the best model performance. Thus, the structure of the overall deep latent neural networks is given as $[I \to 600 \to 200 \to 600 \to I]$, where $I$ is the number of the input data. According to our previous empirical investigations, the activation function of each layer could be either  \textit{softmax} function or \textit{sigmoid} function which largely depends on the cost function $c(\cdot)$ chosen for the calculation of data reconstruction error between $X$ and $X'$. The batch size of the neural networks is set to 500. 
The statistics of the experimental datasets are reported in Table \ref{tab:dataStat}.

\begin{table}[!htbp]
\renewcommand{\arraystretch}{1.2}
    \centering
    \caption{Statistics of experimental data sets.}
    \begin{tabular}{cccc}
    \hline
        &\textbf{ML-20M}&\textbf{Netflix}&\textbf{LASTFM}  \\
         \hline
         \#of users&136,677&463,435&350,200\\
         \#of items& 20,108&17,769&24,600\\
         \#of interactions&10.0M&56.9M&16.1M\\
         \%of interactions&0.36\%&0.69\%&0.16\% \\
         \hline
         \# of held-out users& 10,000&40,000&30,000\\
         \hline
    \end{tabular}
    \label{tab:dataStat}
\end{table}

\subsection{Results on Effect of Control Parameters}

\begin{figure}[!ht]
\centering
\subfigure[ML-20M] {\includegraphics[height=1.1in,width=1.1in]{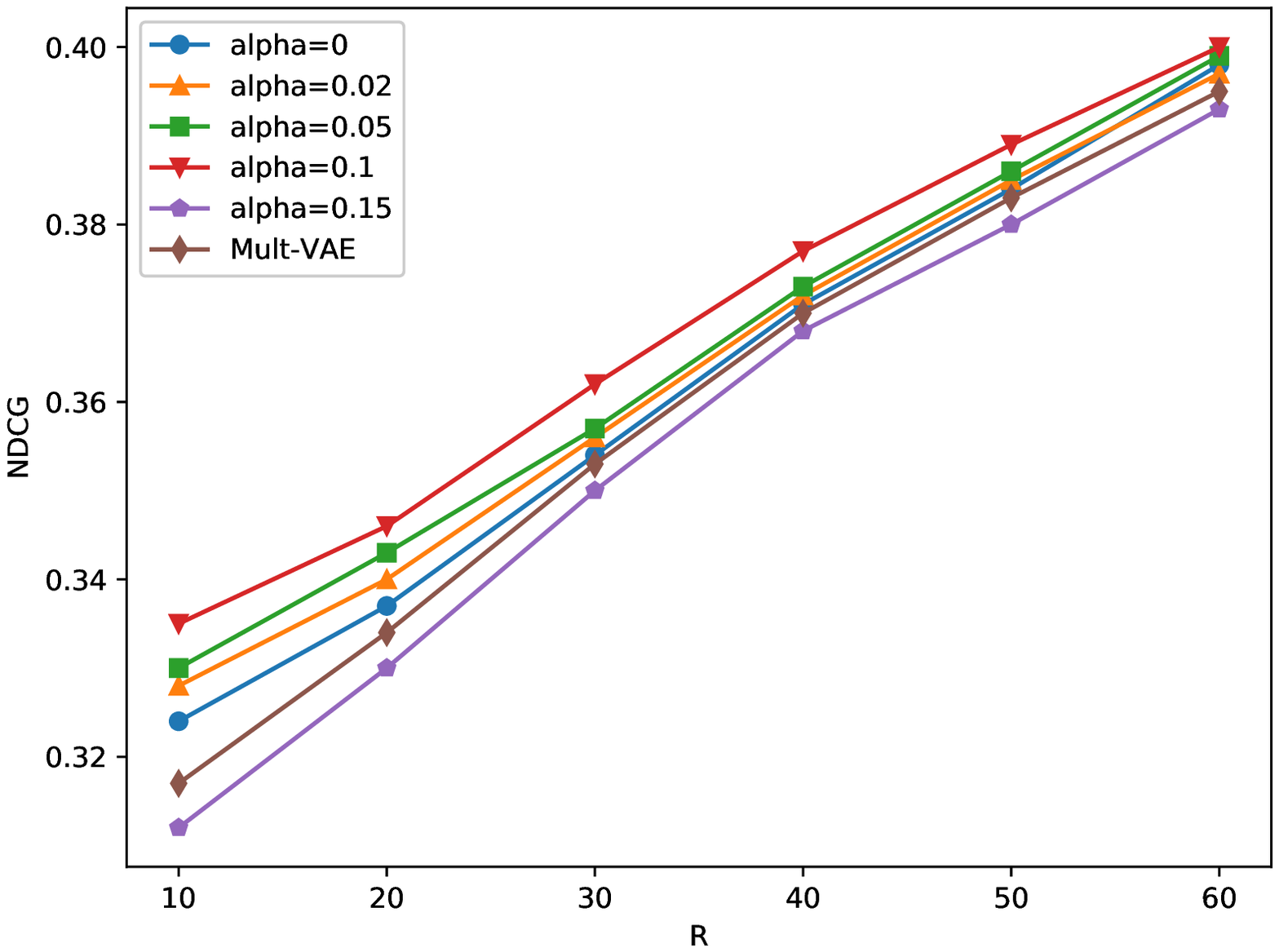}\label{result1}}
\subfigure[Netflix] {\includegraphics[height=1.1in,width=1.1in]{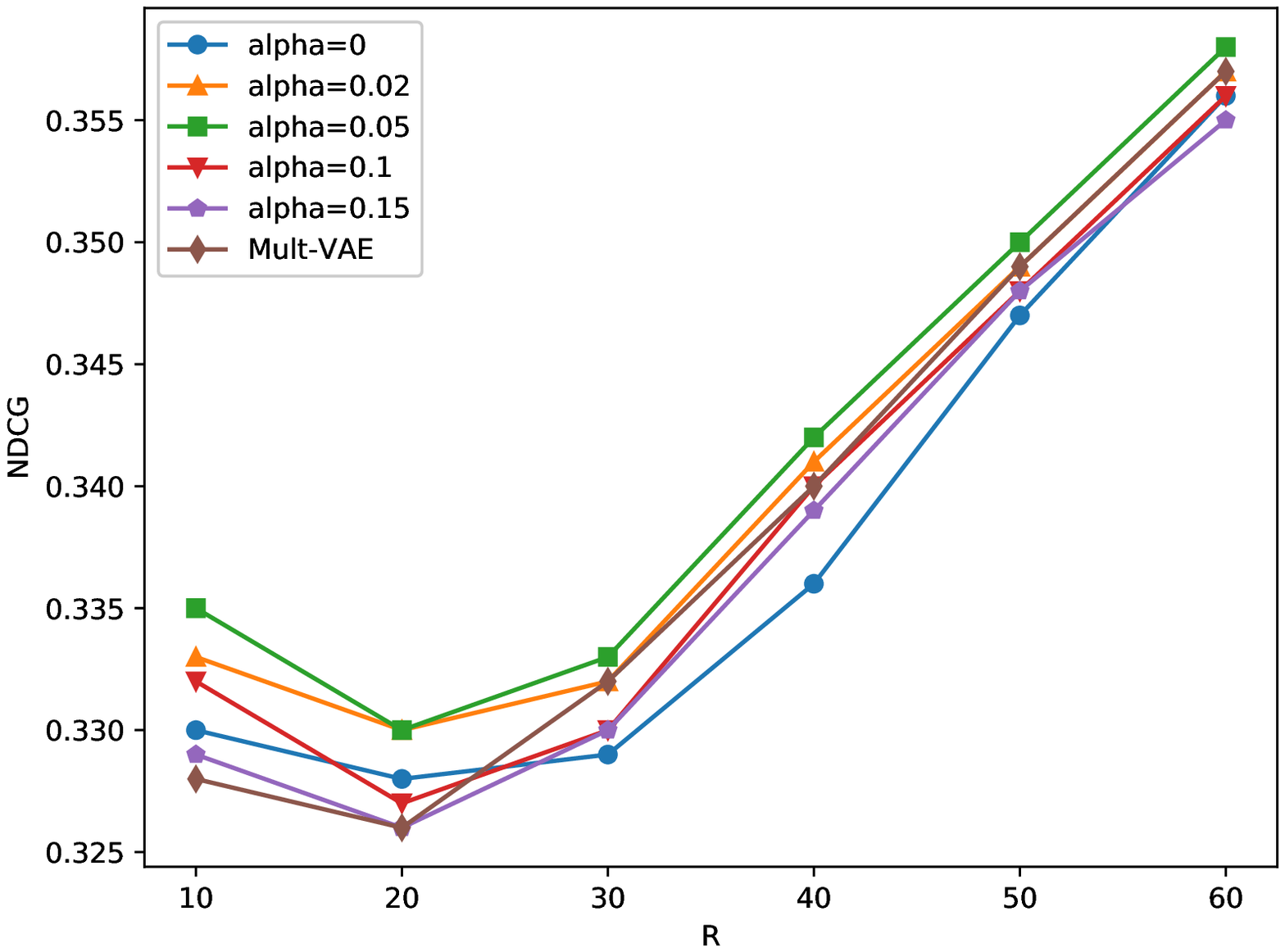}\label{result2}}
\subfigure[LASTFM] {\includegraphics[height=1.1in,width=1.1in]{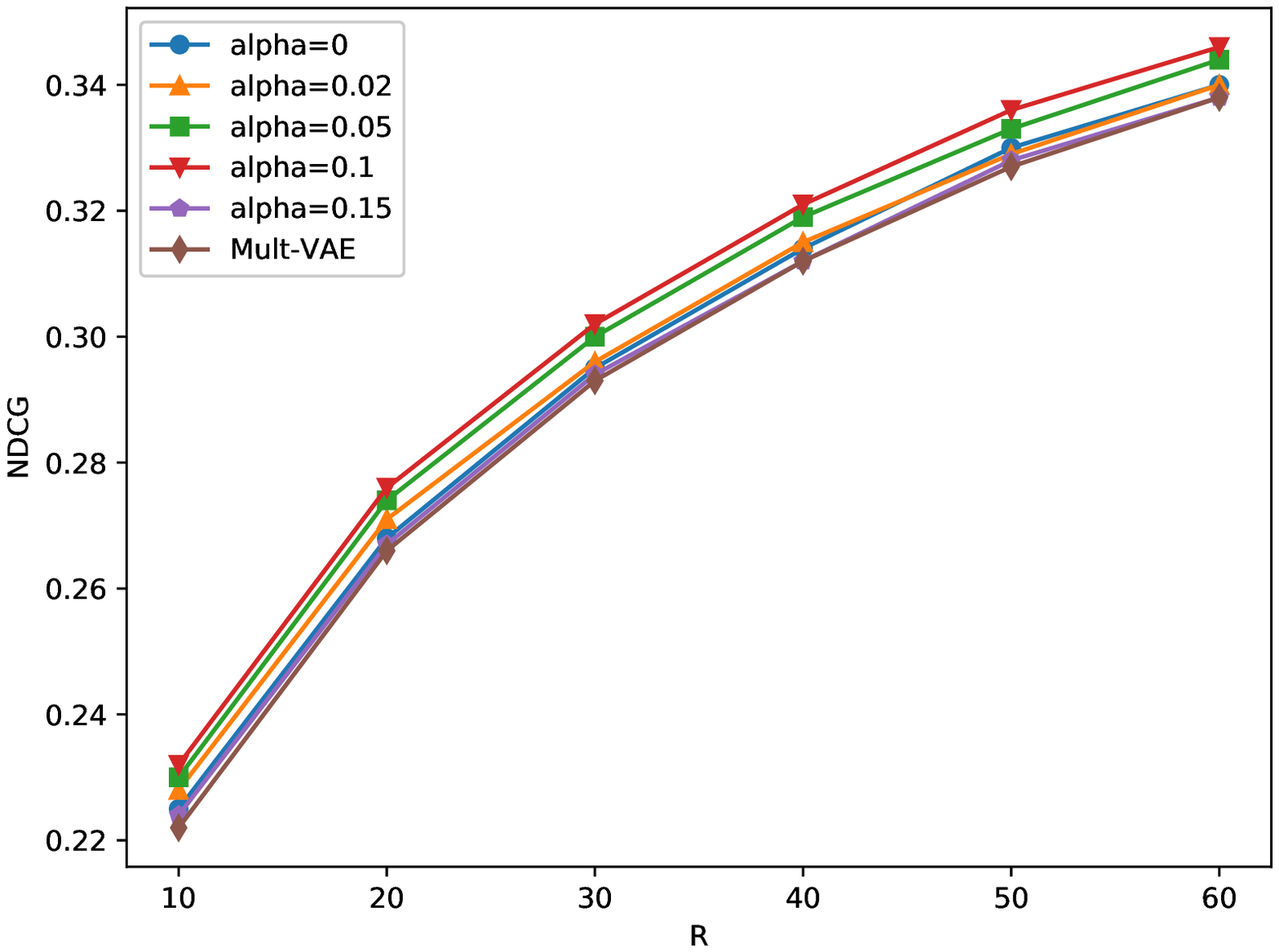}\label{result3}}
\caption{NDCG@R results on three datasets}
\label{NDCGresult}
\end{figure}

\begin{figure}[!ht]
\centering
\subfigure[ML-20M] {\includegraphics[height=1.1in,width=1.1in]{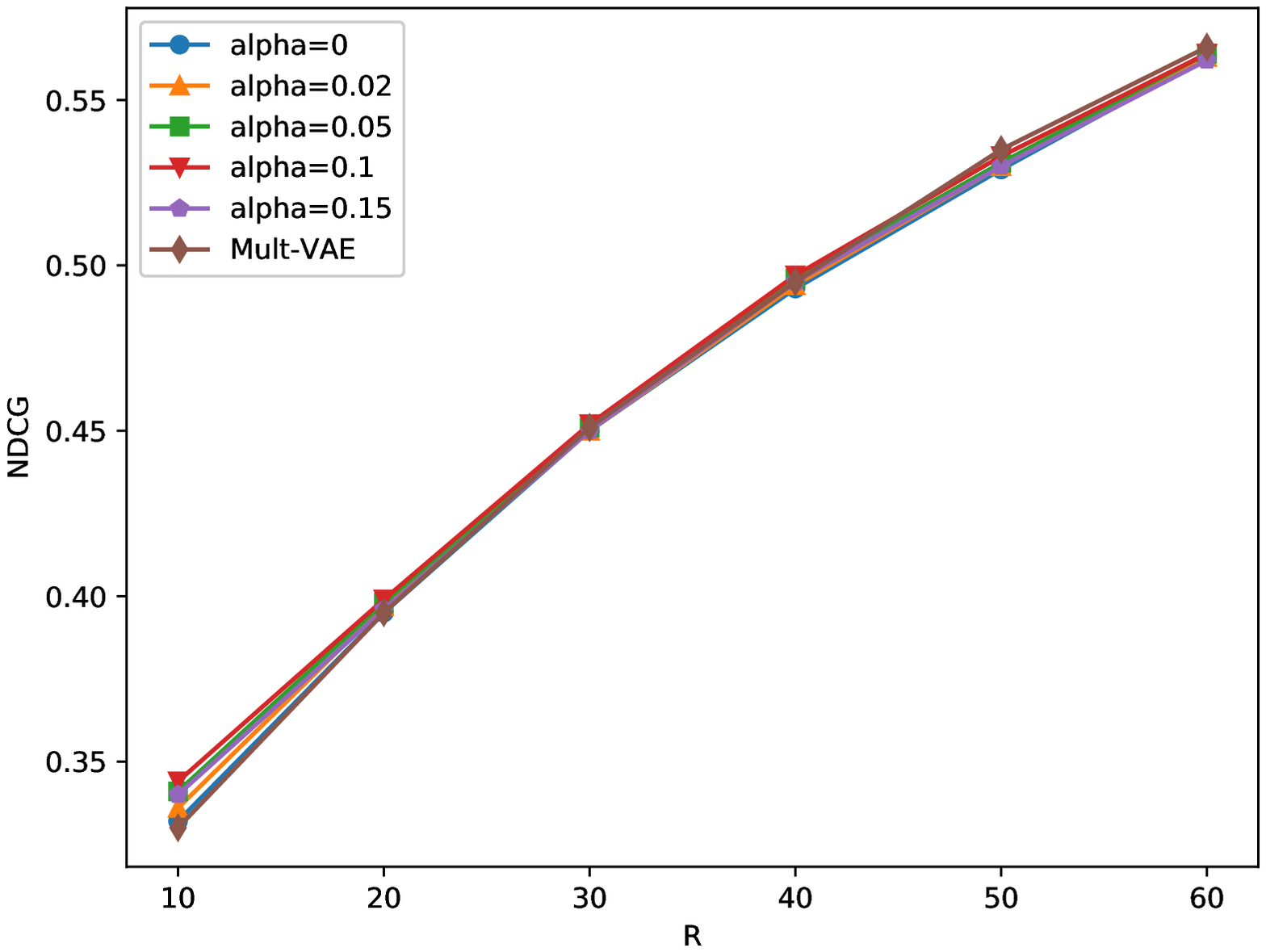}\label{result4}}
\subfigure[Netflix] {\includegraphics[height=1.1in,width=1.1in]{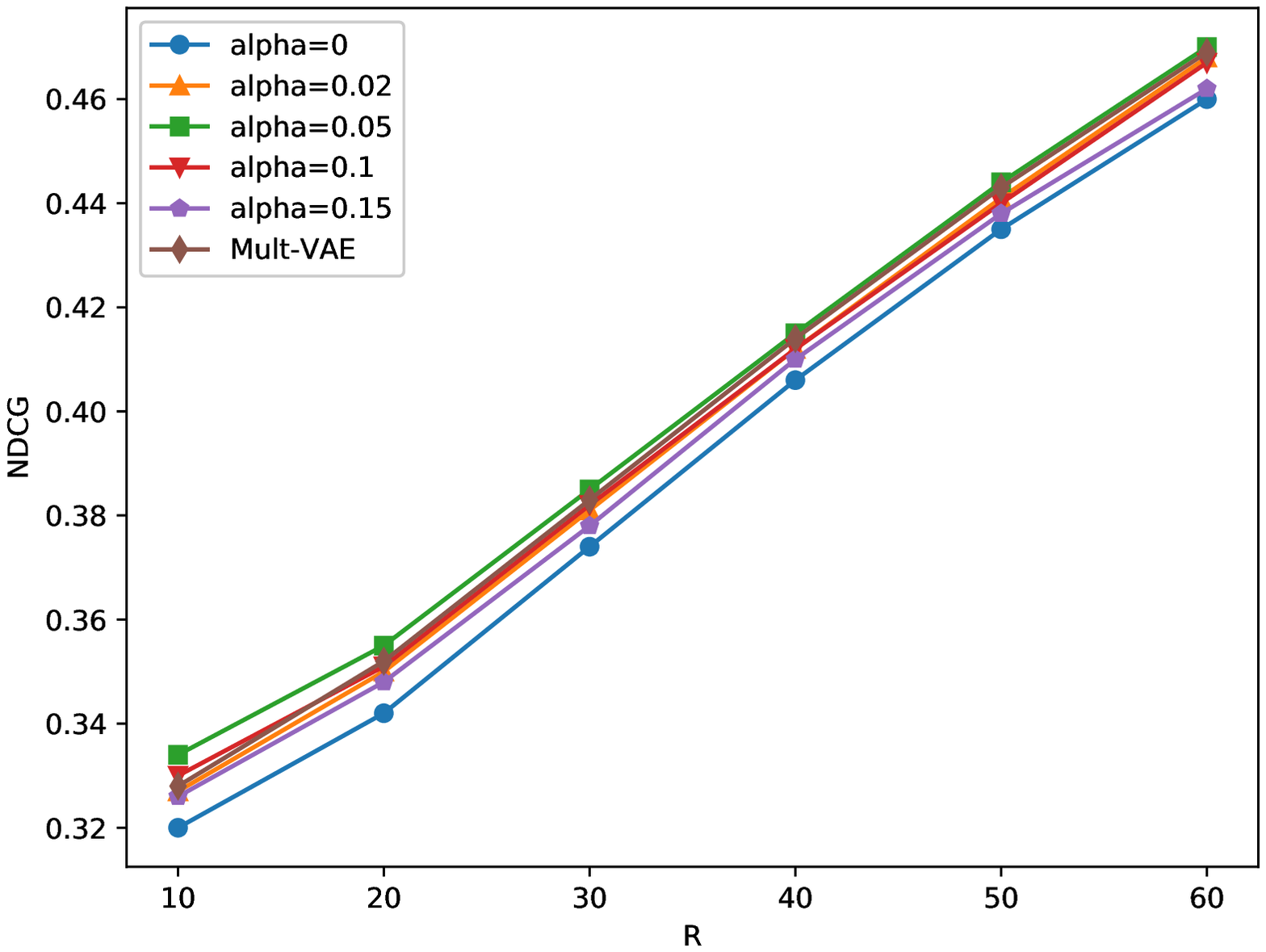}\label{result5}}
\subfigure[LASTFM] {\includegraphics[height=1.1in,width=1.1in]{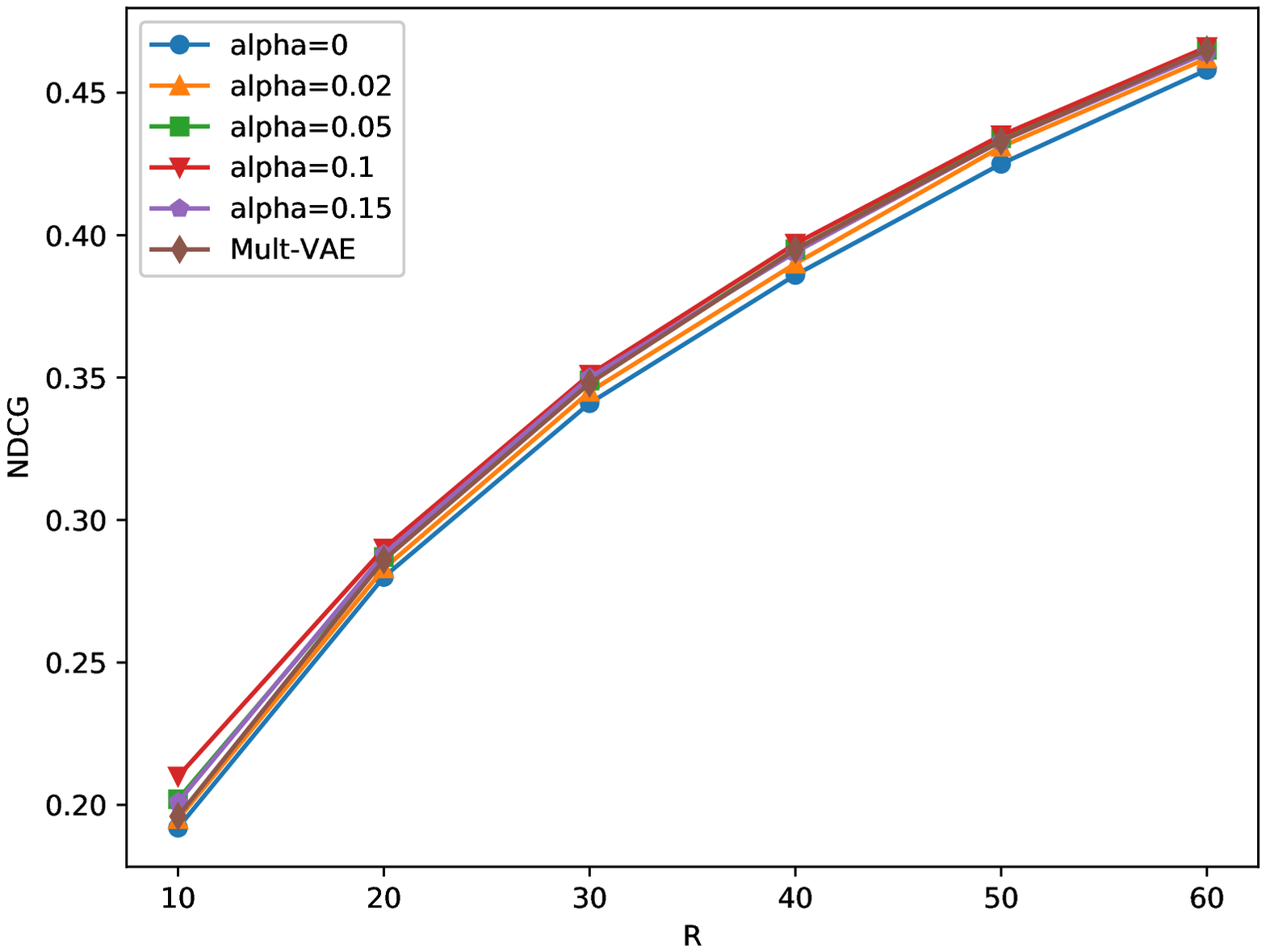}\label{result6}}
\caption{Recall@R results on three datasets}
\label{recallresult}
\end{figure}

We first evaluate how the parameter $\alpha$ can affect the model performance. In this experiment, we set $\alpha=0, 0.02, 0.05, 0.1, 0.15$, respectively, and vary $R$ from 10 to 60 by a step 10. Then, we plot both $NDCG@R$ and $Recall@R$ results of the aWAE and the Mult-VAE on three datasets in Figure~\ref{NDCGresult} and Figure~\ref{recallresult}, respectively.  


From Figure~\ref{NDCGresult}, it is noticed that when $\alpha=0.1$, the model performance ($NDCG@R$ criterion) of the aWAE is the best for ML-20M and LastFM datasets. For Netflix dataset, the model performance of the aWAE is the best when $\alpha=0.05$. And the model performance of most $\alpha$ is better than that of the Mult-VAE which is the state-of-the-art approach. Interestingly, it is well noted that if only a few items are be predicted, e.g., $R\leq 30$, the proposed aWAE is much better than that of the Mult-VAE. This hints that the proposed approach is keen to discover the most interested items for a user, which is very meaningful for many real-world applications. Similar observations could be found in the results on $Recall@R$ criterion, as plotted in Figure~\ref{recallresult}.

\begin{table}[!h]
\renewcommand{\arraystretch}{1.5}
\centering
  \caption{Results on ML-20M}\label{tab:ml}
\begin{tabular}{cccc}
\hline
&Recall@20& Recall@50& NDCG@100\\
\hline
aWAE &	$\textbf{0.400}$&      0.53    &	$\textbf{0.429}$
\\
\hline
Mult-VAE&	0.395      & 	$\textbf{0.537}$ &       0.426
\\
Mult-DAE&	0.387&	0.524&	0.419
\\
\hline
SLIM&0.370 &0.495&0.401\\
CDAE&0.391 &0.523&0.418\\
\hline
\end{tabular}\\

\centering
\caption{Results on Netflix}\label{tab:netflix}
\begin{tabular}{cccc}
\hline
&Recall@20& Recall@50& NDCG@100\\
\hline
aWAE&	0.352&	0.438&	$\textbf{0.386}$
\\
\hline
Mult-VAE&	$\textbf{0.355}$&	$\textbf{0.444}$&	$\textbf{0.386}$
\\
Mult-DAE&	0.344&	0.438&		0.380
\\
\hline
SLIM&	0.347&	0.428&	0.379
\\
CDAE&	0.343&	0.428&	0.376
\\
\hline
\end{tabular}\\

\centering
  \caption{Results on LASTFM}\label{tab:LASTFM}
\begin{tabular}{cccc}
\hline
&Recall@20& Recall@50& NDCG@100\\
\hline
aWAE&	$\textbf{0.287}$&	$\textbf{0.432}$&	$\textbf{0.373}$
\\
\hline
Mult-VAE&	0.279&	0.427 &	0.362
\\
Mult-DAE&	0.277&	0.401&	0.351
\\
\hline
\end{tabular}
\end{table}


\begin{table}[!ht]
\renewcommand{\arraystretch}{1.5}
\centering
  \caption{Results on ML-20M}\label{tab:ml1}
\begin{tabular}{cccc}
\hline
&Recall@1&Recall@5& NDCG@10\\
\hline
aWAE& $\textbf{0.410}$ & $\textbf{0.331}$&	$\textbf{0.333}$
\\
\hline
Mult-VAE&0.378&	0.308&0.318
\\
Mult-DAE& 0.383&0.311&	0.311
\\
\hline
\end{tabular}\\

\centering
  \caption{Results on Netflix}\label{tab:netflix1}
\begin{tabular}{cccc}
\hline
&Recall@1&Recall@5& NDCG@10\\
\hline
aWAE& $\textbf{0.407}$&	$\textbf{0.340}$&	$\textbf{0.335}$
\\
\hline
Mult-VAE& 0.298&	0.287&	0.247
\\
Mult-DAE&0.296&	0.289&	0.248
\\
\hline
\end{tabular}\\

\centering
  \caption{Results on LASTFM}\label{tab:LASTFM1}
  
\begin{tabular}{cccc}
\hline
&Recall@1&Recall@5& NDCG@10\\
\hline
aWAE&$\textbf{0.355}$&$\textbf{0.229}$&	$\textbf{0.228}$
\\
\hline
Mult-VAE&0.317&	0.227&	0.225
\\
Mult-DAE&0.319&	0.228&	0.215
\\
\hline
\end{tabular}
\end{table}
\subsection{Results on Performance Evaluation}
Based on the results of the control parameter experiments, we fix $\alpha=0.1$ 
to perform the rest experiments. We separately evaluate the cases when recommend more items as well as a few items. We report the corresponding comparison results with the rest approaches in the following tables. 

\subsubsection{When recommend more items}
We implement the proposed aWAE and copied the results for all compared models from the original paper for fair comparison, and report the results in Table \ref{tab:ml}, \ref{tab:netflix} and \ref{tab:LASTFM}. From these results, it can be seen that the proposed aWAE achieve the best results on LASTFM dataset, and is better than that of the Mult-DAE and the baseline models, e.g., SLIM and CDAE, on all datasets, but is slightly worse than that of the Mult-VAE on Netflix dataset. 
\subsubsection{When recommend a few items}
In this study, we have implemented the aWAE, the Mult-VAE and the Mult-DAE on all datasets, and the corresponding results are reported in Table \ref{tab:ml1}, \ref{tab:netflix1} and \ref{tab:LASTFM1}, respectively. 

Interestingly, we find that the proposed aWAE is much better than that of the Mult-VAE and the Mult-DAE on all three datasets, especially for the case when only one item is to be recommended. For criterion $Recall@1$, the performance of the aWAE is increased by around $4\%$, $10\%$ and $4\%$ when compared with the Mult-VAE on ML-20M, Netflix and LASTFM dataset, respectively. This further verifies that the proposed approach is meaningful for many real-world applications.

\section{Conclusion}\label{sec:Conclusion}
Conventionally, the collaborative filtering is one of the widely adopted recommendation techniques which is usually applied on the explicit data, e.g., rating scores. Recently, the implicit data including user click data and browsing data is believed to contain user's potential preferences. Therefore, a good number of research efforts have been made towards this end. In the literature, the VAE based approaches have achieved the state-of-the-art performance. However, it might reconstruct multiple input data to the same output data which in turn increases the data reconstruction error. To technically resolve this issue, this paper proposes to adapt the Wasserstein autoencoders for the collaborative filtering task. Our technical contributions are three-fold. First, this is the first attempt to adapt the WAE for the collaborative filtering. Second, we empirically propose the cost function as well as the distance metrics for the sparse implicit data. Rigorous experiments have been evaluated on three widely adopted datasets, i.e., ML-20M, Netflix and LASTFM. Both the state-of-the-art approaches, e.g., Mult-VAE and Mult-DAE, and the baseline models, e.g., SLIM and CDAE, are evaluated and the promising experimental results have demonstrated that the proposed approach is superior to the compared approaches with respect to two widely adopted criteria $Recall@R$ and $NDCG@R$. Interestingly, the proposed aWAE is keen to recommend only few items which is very attractive to many real-world applications.


\bibliographystyle{unsrt}
\bibliography{ijcai19}
\bibliographystyle{ijcai19}

\end{document}